# SOLAR MEAN MAGNETIC FIELD NEAR THE SURFACE AND ITS VARIATION DURING A CYCLE

P.A.P. Nghiem [1], R.A. García [1], and S.J. Jiménez-Reyes [2]

[1] *Service d'Astrophysique, DAPNIA/DSM/CEA CE Saclay*
*AIM – Unité mixte de recherche CEA-CNRS-Université Paris VII, UMR n°7158*
*91191 Gif sur Yvette Cedex, France*
[2] *Instituto de Astrofíca de Canarias, La Laguna, Tenerife, Spain*
e-mail: papnghiem@cea.fr

## ABSTRACT

The unsigned mean magnetic field characterizes the magnetic energy settled in the region near the solar surface. It is one of the determinant ingredients that govern the turbulence in the photosphere, and the magnetic heating of upper layers. The acoustic eigenmodes are directly sensitive to this mean magnetic energy, via the magnetic pressure, and thanks to their trajectories sweeping entirely this region. We use the local-wave formalism to calculate the p-mode frequencies of the Saclay seismic solar model. Then, by comparing them to the LOWL observed frequencies, the $\ell$-independent differences (up to 40 µHz) can be attributed to the existence of a magnetic pressure that modifies the pressure, the density and the sound speed, taking also into account the additional Alfven speed. A profile of unsigned mean magnetic field is deduced, increasing from zero at the surface to $2.5 \times 10^4$ G, 5600 km deeper. Next, by applying the same method, the $\ell$-independent variations in frequency due the solar cycle (up to 0.4 µHz) is used to deduce the change of the magnetic profile. This change presents two distinctive parts: a plateau of only 2-3 G from the surface down to 2100 km, and a very narrow peak of 55 G, 220 km thick, right at the surface. Due to the non-linear effect of the magnetic field, comparing only the frequencies between minimum and maximum activity is not sufficient to deduce the magnetic variation. The additional knowledge of the magnetic field at minimum activity is necessary. If the latter is ignored, an extra variation of up to 130 G would be found. Finally, our results are compared to magnetic field estimates by other helioseismic and spectroscopic methods.

## 1. INTRODUCTION

The surface and sub-surface regions of the Sun are known to be the seat of important magnetic activities. At the beginning, solar physicists focused on active regions because of the presence of intense magnetic fields (kG). Then the field strength between these regions, called the internetwork regions (Lites, Rutten & Kalkofen 1993), have been also the object of numerous measurement campaigns. One of the aims is to evaluate the mean magnetic energy contained in outer layers of the solar plasma, useful for the estimates of the magnetic energy transferred to the corona, and of the convective activity in 3D-solar modelisation. The internetwork region was discovered to be not field free, and although it hosts a weaker field, generally less than 100 G, it covers more than 90% of the surface. It represents the major contributor to the surface magnetism, even at maximum activity (Dominguez Cerdeña, Sánchez Almeida & Kneer 2006).

The mean magnetic energy, or in other words the unsigned mean magnetic field, has been the target of intensive spectroscopic measurement efforts since the beginning of the 90's (see references in section 6). Great difficulties remain to be solved, concerning i) the coherence of local measurements at different heights in the photosphere or the chromosphere, at disk center or limb darkening, ii) the mixed polarity at small scales which tends to cancel the Zeeman signal, iii) the interpretation of the Hanle effect that can only measure low-field strengths above the upper photosphere.

Helioseismic methods should be more suitable, at least on the principle, to estimate the unsigned mean magnetic field, but rather at the base of the photosphere and below. Acoustic waves sweep entirely the sub-surface region, and are directly sensitive to the square field, via the pressure and the Alfven speed. The p-mode frequencies should naturally bear the signature of the mean magnetic energy. Helioseismic estimates of magnetic field have been attempted (see references in section 6). But for various reasons, the frequencies were invoked only for the evaluation of the magnetic increase during the solar cycle. The field at minimum activity was considered as negligible at the surface, while it is evaluated only at deeper layers by studying the frequency splittings. The solar modelisation was reconsidered in order to properly introduce a magnetic field, but generally with a presupposed profile, while the frequency calculation was keeped unchanged, ignoring the magnetic contribution to the sound speed, and using the isothermal atmosphere condition which adds its own correction at the surface.

The present work aims at inferring the unsigned mean magnetic field, and its variation with the solar cycle, by regarding the p-mode frequencies. This will be possible at depths where the magnetic pressure is not negligible compared to the gas pressure. We use partially the results of existing works to account for the magnetic effects on the solar parameters, but acoustic frequencies are calculated with the local-wave formalism, where the solar model is considered as it is, without adding an extra atmosphere, allowing thus the direct study of surface layers. The Alfven speed is also considered when calculating the sound speed.

In the following sections, a summary of the eigenfrequency calculation formalism, along with its results, is given, and the effect of a magnetic field is introduced. Then comparisons between theoretical-observed frequencies at minimum activity will be used to infer the mean magnetic field profile which was not implemented in the initial solar model. Comparisons between observed frequencies at minimum-maximum activities will give the magnetic variation during the solar cycle. Finally, the results will be confronted to other helioseismic and spectroscopic estimates of magnetic field. Implications on radius variations will also be evoked.

We use the Saclay seismic solar model (Turck-Chièze et al. 2001; Couvidat, Turck-Chièze & Kosovichev 2003); the LOWL observed frequencies (Jiménez-Reyes, 2001) of the years 1996-1997, then 1999-2000, resp. for minimum and maximum solar activity.

## 2. EIGENFREQUENCY CALCULATION

In a spherically symmetric star defined by its radial profiles of sound speed $c_0(r)$, density $\rho_0(r)$, pressure scale height $H_p(r)$, the acoustic trapped modes of degree $\ell$ and order $n$, are here characterized by an angular frequency $\omega$ and a phase $\alpha$, which are the solutions of the two following equations

$$\int_{r_1}^{r_2} k_r \, dr = (n+\alpha)\pi \qquad (1)$$

$$\alpha\pi = -\frac{\pi}{2} + \arctan\left\{\frac{1}{x_f^{0.5}}\left[\frac{1}{4x_f} - \frac{Ai'(-x_f)}{Ai(-x_f)}\right]\right\} + \frac{2}{3}x_f^{3/2} \qquad (2)$$

for $\ell \neq 0$

$$\alpha\pi = \frac{\sin(2\pi/a_v)}{(\omega^2/4\pi G\rho_0)+2-\cos(2\pi/a_v)} \qquad \text{for } \ell=0, \quad (3)$$

where the internal turning point $r_1$ is given by

$$k_r = 0 \quad \text{for } \ell \neq 0, \quad k_r = \frac{2\pi}{a_v r} \quad \text{for } \ell = 0 \qquad (4)$$

the external turning point $r_2$ is given by

$$\frac{k^2}{k_r} = \frac{2\pi}{a_v H_p}, \qquad (5)$$

with the radial wave number $k_r$ defined as

$$k_r = \sqrt{\frac{\omega^2 + 4\pi G\rho_0}{c_0^2} - \frac{\ell(\ell+1)}{r^2}}, \qquad (6)$$

and $x_f$ defined as

$$x_f = k_1^{1/3}\left[\frac{(\pi/2a_v)^{2/3}}{2k_1^{1/3}} + \frac{a_v(2\pi)^4}{2(r_2-r_1)^2 r_2^3 k_1^2}\right]. \qquad (7)$$

It remains to precise that
$$a_v = 12.1, \qquad (8)$$

$k_1$ is the derivative of $k_r^2$ at $r_1$, $A_i$ and $A_i'$ are the Airy function and its derivative.

These equations are based on the formalism of local waves, discussed in details in Nghiem (2003, 2006). The main concept was that an acoustic wave can only propagate when, locally, its wavelength is smaller than a characteristic length of surrounding inhomogeneity, e.g. $H_p$, and is reflected in the contrary case. This defines the limits of the resonant cavities which are therefore different for each mode. One of the consequences of this formalism is that the solar model is directly used without any other changes when calculating its eigenfrequencies. One avoids the fitting of the isothermal atmosphere which may complicate the interpretation between predictions and observations. This is important when discussing especially the solar surface region which must include complex phenomena.

The present frequency calculation allows to get rid of this intrinsic problem. Nevertheless, like other semi-analytical calculations, it suffers from the approximations used, and this leads to an imprecision of ± 3 µHz. Another simplification is that only pure acoustic waves can be treated, without any coupling between them or with gravity modes, so that results for low frequencies around 600 µHz should not be valid. But these drawbacks will not trouble the present study.

In Fig. 1, the frequencies obtained are compared to the measured ones of the years 1996-1997, i.e., at minimum activity, for the degrees $\ell = 1$ to 10. The discrepancies has two components. The first component, depending on the degree $\ell$, is the thickness formed by the set of curves of fig. 1: it is the inaccuracy noticed above of ± 3 µHz introduced by the method. The second component, $\ell$-independent, represented by the general trend of these curves, is of greater importance because it reaches -40, +10 µHz. It will actually depend more and more on $\ell$, when $\ell \gtrsim 30$. According to Eq. 6, this very specific behavior clearly shows that these disagreements come from discrepancies at the surface, between the solar model and the Sun.

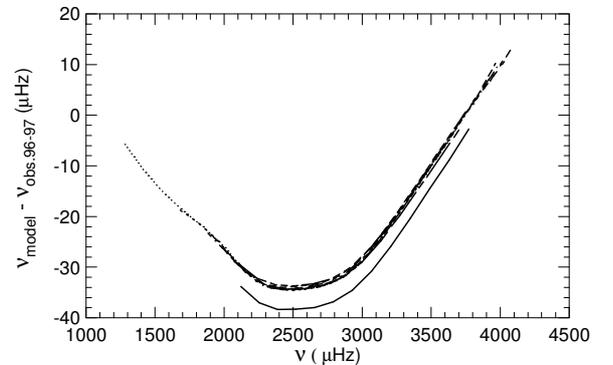

*Fig. 1. Frequency differences, calculations (local-wave formalism) minus observations of the year 1996-1997, for the degrees $\ell =1$ to 10. Points of same $\ell$ are joined together by continuous to progressively discontinuous lines.*

Generally, such a kind of discrepancy is attributed either to non-adiabaticity (Pesnell 1990; Guenther 1994), or else to turbulent pressure at the surface (Rosenthal et al. 1995, 1999). In the first case, with certain parameters describing departure from adiabaticity, the discrepancies are noticeably reduced, without being eliminated. In the second case, a very peaked turbulent-pressure profile is deduced from 3D modelisations, and it allowed to reduce strongly the discrepancies for frequencies above 2500 µHz, but also at the expense of deteriorating the very good agreement at lower frequencies. Note also that these studies were performed under the isothermal atmosphere condition, which will, at each stage, add its own $\ell$-independent modifications.

These attempts mean anyhow that the reason of the theory-observation disagreement should not be searched in a unique direction. In complement to these two alternatives, we propose a third one: the presence of a magnetic field. That will give a greater coherence with the usual assumption of magnetic field variation to account for $\ell$-independent frequency variations with the solar cycle. Here, we will assume that only a magnetic field at the surface and sub-surface regions is at the origin of the calculated-observed frequency discrepancies, and only its variations are at the origin of the solar cycle effects on the observed frequencies. We will thus seek for a magnetic profile that can account for these phenomenons, without searching for the physical mechanism that are at their origin. A work in this direction has been done by Nghiem et al. (2004), but was still incomplete.

### 3. THE MAGNETIC EFFECTS

Let us call $p_0$, $\rho_0$, $c_0$ the pressure, the density, and the sound speed in a magnetic-free gas under hydrostatic equilibrium. The introduction of a magnetic field will lower the gas pressure by a quantity $p_M$, the magnetic pressure which now contributes to maintain the hydrostatic equilibrium. The gas density will also decrease, according to the equation of state. To describe it, we will use the approximated formula given by Li et al. (2003), which has been checked numerically. The sound speed will be modified accordingly, and we have furthermore to add an Alfven component $c_A$. Using the indice $1$ for the parameters in the presence of a magnetic field $B$, we have:

$$p_1 = p_0 - p_M = p_0 - \frac{B^2}{8\pi} \qquad (9)$$

$$\rho_1 = \frac{\rho_0}{1 + \frac{p_M}{p_0}} = \frac{\rho_0}{1 + \frac{1}{p_0}\frac{B^2}{8\pi}} \qquad (10)$$

$$c_1^2 = \frac{\Gamma_1 p_1}{\rho_1} + c_A^2 = c_0^2 + \frac{1}{\rho_0 p_0}\frac{B^2}{8\pi}\left[2p_0 + (2-\Gamma_1)\frac{B^2}{8\pi}\right] \qquad (11)$$

where $\Gamma_1$ is the adiabatic coefficient that is supposed here to be invariant.

In fact, these equations should be a little more complicated. The expression of $p_M$ in Eq. 9 and 10 should be accompanied by a coefficient $\beta$ that depends on the direction of the magnetic field relatively to the horizontal one. In Eq. 11, $c_1^2$ should further depend on a term in $\cos^2\theta\, c_0^2 c_A^2$, where $\theta$ is the angle between the magnetic direction and the propagation direction. To simplify, we have searched only for a horizontal magnetic field, so that $\beta = 1$, and as at the surface, the low-degree modes considered here will be vertical, $\cos\theta = 0$. We should use $B_h$ instead of $B$, but we have omitted the indice $h$ for the sake of simplicity. On another side, very few is known about the direction of the magnetic field. It would be possible that the magnetic energy is accumulated at the base of the convective zone as a toroidal field, until it rises up to the surface by keeping its horizontal direction, except in active regions where there is a very strong concentration of vertical field lines. Photospheric measurements done by Lites et al. (1996, 1998) at emergence sites seem to be compatible with this scenario. Coherent with that are the measurements on the Na profile by García et al. (1999), giving a very weak mean magnetic field of 0.12 G in the line of sight direction. If the mean magnetic field is actually predominantly horizontal, our simplification to $B_h$ would not be very restrictive.

It is also predictable that at certain location, the solution found $B^2$ can be negative, the gaseous pressure and density have to be increased instead of decreased. This means that the frequency disagreement with observation cannot be attributed to a magnetic field there, but is due to a displacement of the gas from magnetic regions toward this region, as the total mass must be conserved. In this case, the modified physical parameters are expressed as:

$$p_1 > p_0 \qquad (12)$$

$$\rho_1 = \frac{\rho_0}{2 - \frac{p_1}{p_0}} \qquad (13)$$

$$c_1^2 = \frac{\Gamma_1 p_1}{\rho_1} = c_0^2 - \frac{\Gamma_1}{\rho_0 p_0}(p_1 - p_0)^2 \qquad (14)$$

Suppose now that the field $B$ increases by a quantity $\delta B$, the new physical parameters, with the indice $2$, become, when expressing with the former ones of indice $1$:

$$p_2 = p_1 - \frac{M}{8\pi} \qquad (15)$$

$$\rho_2 = \frac{\rho_1}{1 + \dfrac{M}{8\pi p_1 + 2B^2}} \qquad (16)$$

$$c_2^2 = c_1^2 + \frac{M/8\pi}{\rho_1\left(p_1 + \dfrac{2B^2}{8\pi}\right)}\left[2\left(p_1 + \frac{B^2}{8\pi}\right) + (2-\Gamma_1)\left(\frac{2B^2 + M}{8\pi}\right)\right] \qquad (17)$$

where

$$M = 2B\delta B + (\delta B)^2 \qquad (18)$$

It can be seen in the equations (9) to (18) that by comparing between the states 1 and 0, i.e. with magnetic field (the Sun) and without magnetic field (the solar model), the magnetic field $B$ can be inferred. On the contrary, a comparison between the states 2 and 1, i.e. between maximum and minimum activity, is not sufficient to infer the field variation $\delta B$. As the magnetic effect is not linear, because of its effect in $B^2$ and also of its non linear effect on the density, the prior knowing of $B$ is necessary to know $\delta B$. It is important to keep that in mind.

## 4. RESULTS: THE MEAN MAGNETIC FIELD PROFILE

Using Eq. (9) to (11) to modify the solar model, we search for the profile of $B$ that can cancel the $\ell$-independent frequency differences in Fig. 1. This procedure is based on the property that each frequency has its own external turning point $r_2$ which is independent of $\ell$, for the low $\ell$ considered here, and of course a modification outside a cavity will not affect the corresponding frequency. To allow the reader to be able to distinguish the consequences due to different frequencies, we give in Table 1 the turning points corresponding to some frequencies. This table gives also the limited radius range that can be studied, due to the limited number of frequencies.

Table 1. External turning points corresponding to some frquencies, in relative radius units, with the convention that $r/R = 1$ is the photospheric radius.

| $\nu$ (µHz) | $r_2 / R$ |
|---|---|
| 1102.0 | 0.98757 |
| 1546.1 | 0.99320 |
| 2099.8 | 0.99702 |
| 2508.1 | 0.99849 |
| 3052.4 | 0.99955 |
| 3325.0 | 0.99986 |
| 3598.7 | 1.00007 |
| 3873.4 | 1.00021 |
| 4010.5 | 1.00027 |

Following Fig. 1, we can estimate that the disagreement with observation is zero at $\nu \sim 1000$ µHz. We start from $B = 0$ for $\nu \sim 1000$ µHz, i.e., $r/R \sim 0.988$, then search for $B$ at larger radius in order to cancel the disagreements for successively larger frequencies. That means that an environment modification for a given frequency will depend on the results already obtained for lower frequencies, but will not change them. The resulting profile of $B$ is given by the continuous line in Fig. 2, and the corresponding change in sound speed, pressure, and density are given in Fig. 3.

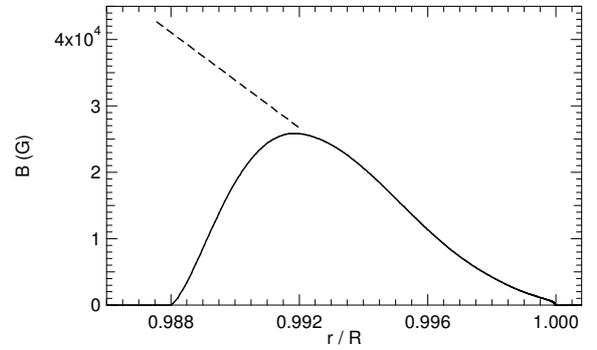

*Fig. 2. Profile of the unsigned mean magnetic field that cancels the $\ell$-independent differences between calculated and observed frequencies (continuous line). The dashed line indicates a more plausible increase of the magnetic profile downward.*

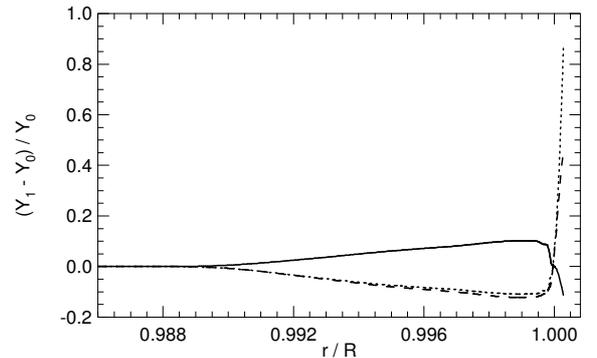

*Fig. 3. Relative change of the sound speed (continuous), the pressure (dashed) and the density (dotted), due to the magnetic profile of Fig. 2.*

Several important remarks must be made:
- As the magnetic effect on the considered physical parameters is in $B^2$, and as the p modes cover all the azimuthal directions, $B$ must be understood as the square root of an average over the azimuths of $B^2$, or in short, the unsigned mean magnetic field.

- When varying the initial frequency and thus the initial radius where B=0, the first increasing part of the profile will change, but it appears that the maximum of 25 kG is reached anyway at $r/R \sim 0.992$, corresponding to ~1500 µHz, and to ~5600 km below the photosphere, and then the decreasing part of the profile is always the same.
- There is actually no physical reason for B to drop to zero when going donward. The result found here is probably due to the fact that the magnetic pressure, although continuing to increase downward, becomes nevertheless negligible at a certain depth compared to the gas pressure that increases much more rapidly. A more plausible increase of B toward the interior is represented by the dashed line in the same figure.
- It appears that B drops to zero very near the photospheric radius, and this relatively quickly, because it still amounts to 1 kG at 300 km deeper.
- Above the photospheric radius, the solution $B^2$ is negative, thus non physics, and Eq. (12) to (14) are used instead. This can be interpreted as due to the conservation of the total mass. The missed gas in the less dense magnetic regions, has been displaced to outer regions that become denser (Fig. 3). This can also be regarded as a change of the radius containing the same mass. When compared to solar models where this kind of radius is used, the actual radius would be larger in the regions inside the photospheric radius, and smaller outside.

With the modifications on the solar model as indicated in Fig. 3, due to the magnetic profile of Fig. 2, the final frequency disagreements obtained are shown in Fig. 4. There remains only differences that depend on $\ell$, which are at the most ± 3 µHz, due to the imprecisions of the frequency calculation method. An extra detail should be precised. The LOWL frequencies of radial modes ($\ell = 0$) are not available in the same conditions than here. If we use the GOLF data (García et al. 2001) instead, the final disagreement will be a line in Fig. 4, between that of $\ell = 1$ and those of other $\ell$. For the sake of coherence between observed frequencies, we do not include this result here.

## 5. RESULTS: THE MAGNETIC VARIATION WITH SOLAR ACTIVITY

Since Woodard and Noyes (1985), it is well established that during the solar cycle, the acoustic frequencies vary, in a way that looks like the disagreement between calculated and observed frequencies, i.e., independently of $\ell$ for low $\ell$, and depending more and more on $\ell$ for $\ell \gtrsim 30$. It is thus natural to think in the framework of the present study that this variation is due to a variation of the mean magnetic field in the sub-surface. This is anyway an assumption already made by numerous studies (see section 6).

Fig. 5 gives the LOWL observed frequencies at minimum activity minus those at maximum activity, for $\ell = 1$ to 10. Despite small variations around a general $\ell$-independent tendency, the latter can be seen easily, and is materialised in the figure by the big dark points, which mark the values that are asked to the frequency calculation to account for.

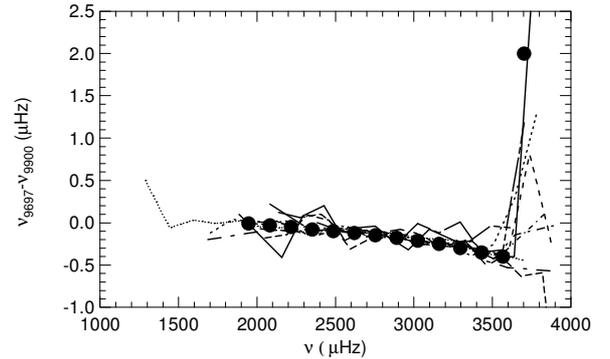

*Fig. 5. Frequency differences, years 1996-1997 minus years 1999-2000, for the degrees $\ell = 1$ to 10. Points of same $\ell$ are joined together by continuous to progressively discontinuous lines. The big dark points are the values that are asked to the frequency calculation to account for.*

We employ exactly the same procedure as in the previous section, but with Eq. (15) to (18). Assuming that the effect of the activity cycle begins from the frequency 2000 µHz, we start the calculations from a zero field variation at $r/R = 0.9966$. The resulting profile of magnetic variation $\delta B$ is given in Fig. 6, leading to the variations in sound speed, pressure and density given in Fig. 7. Like in the previous section, take note that $\delta B$ could probably drop to zero much deeper than what is found here, but the magnetic-pressure change becomes too weak compared to the gas pressure that increases strongly toward interior, and its effect becomes undetectable.

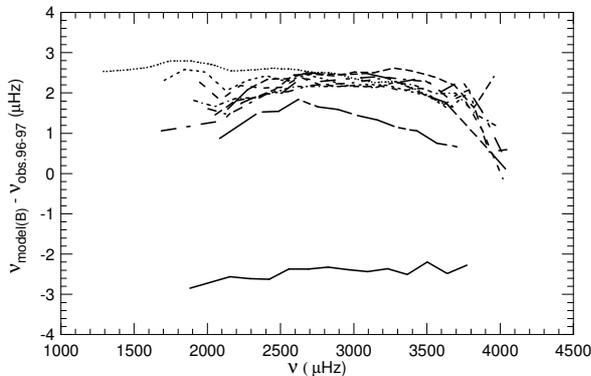

*Fig. 4. Frequency differences, calculations with the profiles modified as in Fig. 3, minus observations of the year 1996-1997, for the degrees $\ell = 1$ to 10. Points of same $\ell$ are joined together by continuous to progressively discontinuous lines.*

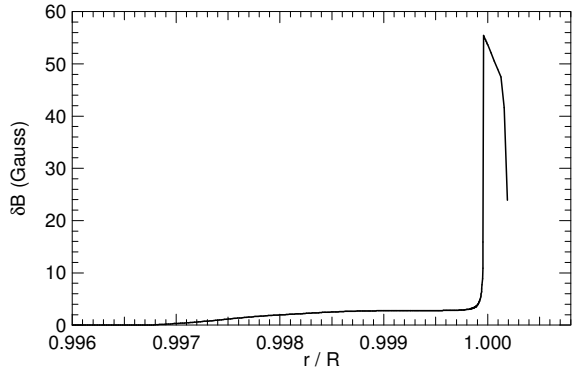

*Fig. 6. Profile of the magnetic field variation to account for the ℓ-independent frequency variation with the solar cycle.*

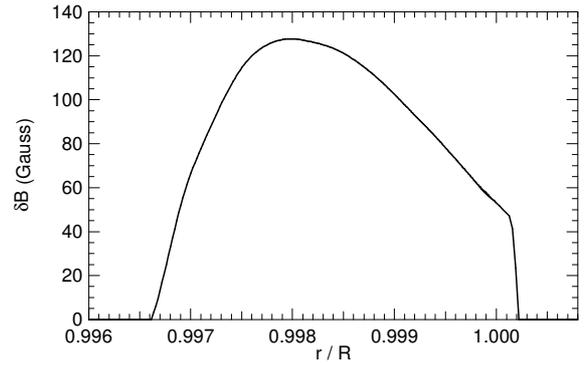

*Fig. 8. Profile of the magnetic field variation to account for the ℓ-independent frequency variation with the solar cycle, but ignoring the mean field B.*

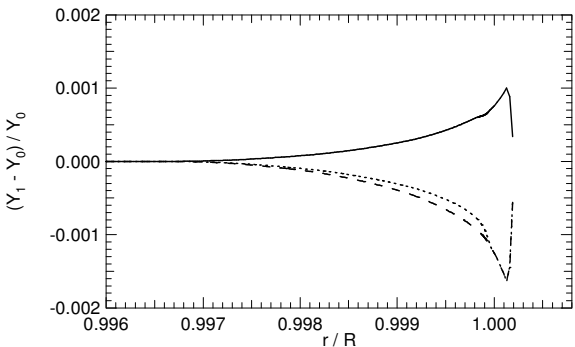

*Fig. 7. Relative change of the sound speed (continuous), the pressure (dashed) and the density (dotted), due to the magnetic profile of Fig. 6*

The most remarkable point is the two distinctive parts of the δB profile: 3 G lying on a wide region from 70 to 2360 km depth, then 55 G in a very narrow region of 220 km thick around the photospheric radius.

Toward the exterior, the abrupt decrease of δB is dictated by the frequencies from 3500 to 3900 μHz. This is at the origin of the violent frequency variations during the cycle, observed for frequencies beyond 3500 μHz (e.g. Jiménez-Reyes et al. 2001, Fig. 5). Those variations are systematically observed, but untill now their reality is still questionable. They are quite understandable in the frame of the local-wave formalism combined with a rapid magnetic field decrease.

Toward the interior, the abrupt transition of δB from 55 to 3 G is due to the abrupt decrease to zero of the mean field *B* (see Fig. 2). Indeed, at these radius, the frequency variations are smooth, and the induced parameter variations of Fig. 7 are also smooth. If the mean field *B* is ignored, we will arrive to a profile of δB with the same external part but without any abrupt transition inside (Fig. 8). In this case, the maximum of δB will seem to be much larger, and situated much deeper.

## 6. OTHER MEAN MAGNETIC FIELD ESTIMATES

Very few observation-based estimates of the mean field below the photosphere are available until now. Goode & Dziembowski (1993) have put an upper limit of 1 MG at the base of the convective zone when studying the mode splittings. Basu (1997) using splittings of modes with their lower turning points around this region, gave an upper limit of 0.3 MG. Antia, Chitre & Thompson (2000) got a similar upper value, 0.3-0.4 MG, and noted that it could become larger by a factor $\sqrt{2}$, if the thickness of this region is reduced by a factor 2. They also found a possible presence of a 20 kG field around the radius *r/R* = 0.96. Li & Sofia (2001) started from this last result and found that this value can increase up to 47 kG at maximum activity. These peak values are obtained with an assumed Gaussian shape having a given width.

Our results cover a much shallower region. In order to have a crude comparison with these studies, we can make a linear extrapolation of our magnetic profile toward interior, like indicated in Fig. 2, although there is no reason for such a linear behaviour to persist on such a large distance. It can be found that the fields at *r/R*=0.7 and 0.96 would be respectively 1 MG and 140 kG.

The magnetic field variation during the solar cycle in near surface regions, has been more studied, using helioseismic results. Goldreich et al. (1991) compared the acoustic frequencies between maximum and minimum activity, to deduce a field variation of 250 G at *r/R*=1, increasing regularly toward interior, up to 800-2300 G at *r/R*=0.995-0.922. These values have been confirmed later by Dziembowski, Goode & Schou (2001). These studies take into account the changes on the entropy, the path length and the sound speed, including the Alfven speed. But Kuhn (1998) disputed these too large variation amplitudes, pointing that MDI magnetograms exhibit a variation of only 67 G at the surface. Other important efforts have been done, by Lydon & Sofia (1995) and Li et al. (2003), to thoroughly explored magnetic effects and turbulence effects, and then introduced them into solar modeling. That leads to a profile of magnetic change of only 45 G up to 150 G, from the surface down to the radius $r/R= 0.9966$, when searching to reproduce cyclic changes of acoustic frequencies, solar irradiance and photospheric temperature.

Our result, $\delta B = 55$ G right at the photospheric radius, is very close to that of Kuhn (1998) and Li et al. (2003). On the contrary, our deeper profile of $\delta B$ is only similar to that of the last authors when ignoring the mean field B. This suggests that this $\delta B$ should be much lowered toward the interior when the mean field $B$ is taken into account.

It is also interesting to notice that the resulting decrease of the density (Fig. 7) in the region immediately below the surface, means that the density scale height $H_\rho$ has decreased. Therefore the f-mode radius which are defined by a given value of $H_\rho$, will also decrease. This is coherent with what has been found by Lefebvre & Kosovichev (2005), and Lefebvre et al. (2006) who studied the cycle effect on radius associated to f-modes, and found that they decrease, in the region between the surface down to $r/R = 0.99$, reaching the maximum at $r/R = 0.9955$.

On another hand, many spectroscopic observations have been performed to measure the mean unsigned magnetic field in the photosphere. Based either on the Zeeman effect or the Hanle effect, these observations aim to explore exclusively the weak magnetic field of internetwork regions, discarding the kG field of active regions. Faurobert-Scholl et al. (1995) indicated a mean field of 20-10 G at photospheric heights 200-400 km, which becomes at maximum activity 30-20 G (Faurobert et al. 2001). Domínguez Cerdeña, F. Kneer & Sánchez Almeida (2003), and Sánchez Almeida (2003) put forward an average field of 15-20 G, which would not vary more than 40% with the activity cycle. But more recent estimates seem to show a much greater dispersion. Lites & Socas-Navarro (2004) found a mean field of only 8 G with high-resolution spectrometric observations, while Trujillo Bueno, Shchukina & Asensio Ramos (2004) report a field of 130 G. Khomenko et al. (2005) combined infrared and visible observations to arrive to an average of 20 G, while Dominguez Cerdeña, Sánchez Almeida & Kneer (2006) gathered observations of different sources, inluding kG field measurements, to build a probability density function, and arrived to an average field of 150 G.

Although these spectrocopic estimates are still to be refined, and our results concern rather regions at the base of the photosphere and lower, they seem not to be strongly incompatible. Our results show a mean field varying during the cycle between 0 and 55 G at $r/R = 1$, averaging over any weak or strong field regions. These values are in the same range as an internetwork field of some tens of G. This strong variation associated with a weak variation of only 3 G in layers immediately below, could indicate that during the cycle, the internetwork field vary only moderately, and most variations come essentially from active regions concentrated right at the base of the photosphere.

## 7. DISCUSSIONS

We have inferred a profile at the surface and sub-surface layers of the unsigned mean magnetic field $B$ and its variation $\delta B$ during the solar cycle. The mean field is proposed as a third alternative, following non-adiabaticity and turbulent pressure, to explain the frequency differences between theory and observations. It has the merit to be coherent with the usual explanation of magnetic field variation to account for the frequency variations during the activity cycle, which is of the same nature, i.e. mostly independent of $\ell$. Furthermore, due to the non-linearity of magnetic effects, it is necessary to obtain $B$ before to calculate $\delta B$. With the present point of view, a more coherent scheme can also be outlined for the magnetic energy distribution according to the traditional dynamo theory: magnetic energy is strongly concentrated at the base of the convective zone, until a threshold is exceeded, then it floats toward the surface while being progressively less strong under the smearing effect of convective motion. The mean magnetic field can thus be expected to regularly decrease from the base of the convective zone toward the surface. The profile of $\delta B$ presents two distinctive parts: a strong narrow peak around the photospheric radius, and a weak plateau deeper. That would mean that at maximum activity, the magnetic energy arriving to the surface is only slightly higher, but when reaching the photosphere where the gas pressure is the lowest, the magnetic field is less affected by the convective motion of the gas, and can be restructured to become stronger. In other words, a small amount of magnetic energy could be shortly accumulated right at the surface, before to be destroyed by specific gas circulations there.

The presence of magnetic energy is expressed here in the form of structural changes: the pressure and the density decrease, the sound speed increases due to the addition of the Alfven speed. There is thus less matter where the magnetic energy is present, and as the total mass must be conserved, the matter is displaced to regions without magnetic energy. This implies inhomogeneous changes in radius, in a way that depends on the presence or absence of magnetic field, and also on the kind of radius in question. When the radius containing the same mass is considered, it expands where there is a magnetic field, and shrinks where the latter drops to zero. When the radius of f-modes is considered, it varies in the opposite sense.

The results presented here could be useful as guidelines to study the competing forces of convection and magnetism along the convection zone. They attempt also to fill the gap between magnetic estimates at the base of the convective zone by helioseimic observations, and that of spectrometric observations in the high photosphere. But much remain to do, in order to make all these different results perfectly coherent, and to find the physical mechanism capable of describing properly these observations.

**Acknowledgements.** We are indebted to Sylvaine Turck-Chièze for her useful suggestions.